\documentclass[twocolumn,superscriptaddress,showpacs,10pt,prl,floatfix]{revtex4}

\usepackage{graphics}%
\usepackage{amssymb}

\begin{document}

\title{Conductance fluctuations at the quantum Hall plateau transition}

\author{F.~Hohls}
\email{hohls@nano.uni-hannover.de}
\author{U.~Zeitler}%
\author{R.~J.~Haug}
\affiliation{%
Institut f\"ur Festk\"orperphysik, Universit\"at Hannover, Appelstr.~2, 30167
Hannover, Germany }

\date{\today}

\begin{abstract}

We analyze the conductance fluctuations observed in the quantum Hall regime for
a bulk two-dimensional electron system in a Corbino geometry. We find that
characteristics like the power spectral density and the temperature dependence
agree well with simple expectations for universal conductance fluctuations in
metals, while the observed amplitude is reduced. In addition, the dephasing
length $L_\Phi \propto T^{-1/2}$, which governs the temperature dependence of
the fluctuations, is surprisingly different from the scaling length
$L_{sc}\propto T^{-1}$ governing the width of the quantum Hall plateau
transition.

\end{abstract}


\maketitle

\newlength{\plotwidth}          
\setlength{\plotwidth}{7.5cm}

\newcommand{\rms}{\mathrm{rms}}

Conductance fluctuations are probably one of the most prominent features
occurring in the mesoscopic world. For metallic (diffusive) systems at low
enough temperature these so-called universal conductance fluctuations (UCF) are
characterized by an amplitude $\sim e^2/h$, independent on sample  size and
material, and, they are generally believed to be well understood~\cite{lee87}.

Similar conductance fluctuations are also observed in mesoscopic quantum Hall
(QH) systems, however, the underlying physics still remains an open question.
Interpretations of different experiments cover modifications of UCF in high
fields~\cite{timp87,geim92,hohls99}, tunnelling between opposite edge states
through bulk inhomogeneities~\cite{simmons91,main94,bykov96}, influences of
charging effects~\cite{cobden96,cobden99}, and networks of
compressible-incompressible regions~\cite{machida01}. Common to all the
previous experiments is a sample geometry in the form of a (Hall-)bar, the
importance of edge channels, and a typical sample width $w$ of 1-3~$\mu$m.
Transport in edge channels and the coupling of both edges due to single
impurities is strongly relevant, making the applicability of a general
UCF-theory problematic.

Here we will address the conductance fluctuations in the quantum Hall regime
due to transport through the disordered bulk of a two-dimensional electron
system (2DES). In order to avoid any edge effects we use a Corbino geometry.
The large width $w=6\,\mu$m, considerably above the elastic scattering length
$l_{\rm el} < 1\,\mu$m, suppresses transport through individual impurities.
Analy\-zing the temperature dependence of small fluctuations superimposed on
the conductance peak of the QH plateau transition we find a behavior very
common to UCF in metallic systems, however, with a considerably reduced
absolute value of the UCF amplitude.

The samples used for this work are based on modulation doped GaAs/AlGaAs
heterostructures. Additional impurities (Be or Si) added into the
2DES~\cite{ploog87,haug87} yield an enhanced short range scattering and a small
elastic scattering length. Throughout this paper we present data measured on a
sample with an electron mobility $\mu=2$~m$^2$/Vs, an electron density
$n=2.1\cdot 10^{15}$~m$^{-2}$ and a density of Be-impurities  $n_{Be} = 2\cdot
10^{14}$~m$^{-2}$. We confirmed our results with measurements on other samples,
containing both Si- and Be-impurities, with mobilities ranging up to
$\mu=12$~m$^2$/Vs. Using annealed AuGe-Ni contacts the samples were patterned
into Corbino geometry with a ring width $w=6\,\mu$m, and an inner radius
$r=60\,\mu$m.

The samples were mounted on the cold finger of a dilution refrigerator with a
base temperature $T < 20$~mK and positioned into the center of a
superconducting solenoid. We measured the current with a Lock-In technique for
fixed AC voltage (9~Hz) as a function of magnetic field and temperature. We
carefully checked for heating effects by varying the amplitude of the
excitation voltage at the same temperature. Excitation amplitudes adapted to
the different temperature regimes were used to avoid heating at the lowest
temperatures and still ensure high enough resolution of the fluctuations at
elevated temperatures. The voltage was fixed to $1\,\mu$V for $T < 50$~mK,
$2\,\mu$V for $50 \leq T \leq 120$~mK, and $5\,\mu$V for $T > 120$~mK.

\begin{figure}[b]
  \begin{center}
  \resizebox{\plotwidth}{!}{\includegraphics{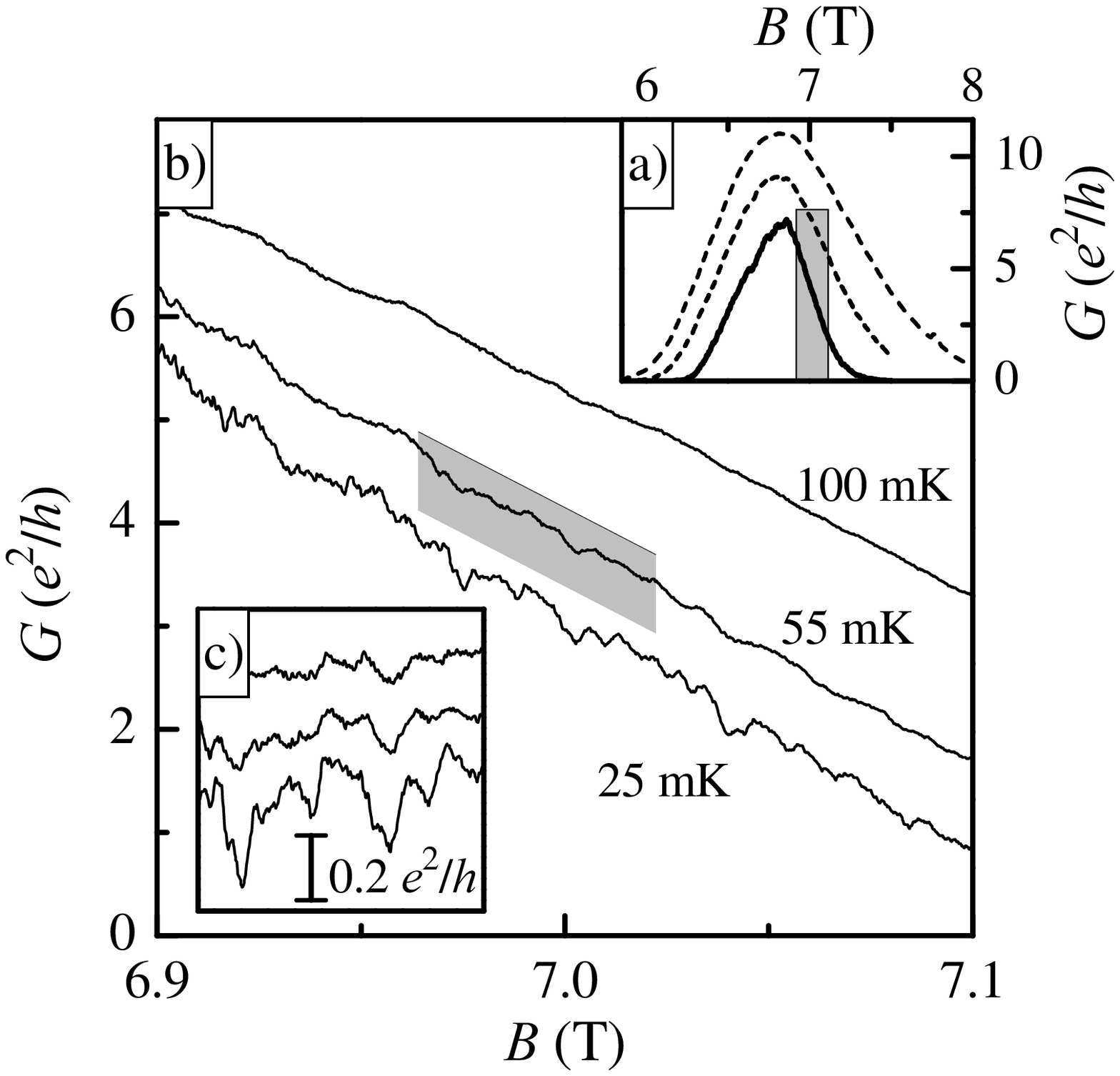}}
  \end{center}
  \caption{a) Conductance peak at the
    $\nu=2\rightarrow 1$ plateau transition for temperatures
     $T=55$~mK (solid line), 160~mK, and 320~mK. \newline
     b) Enlargement of the grey area in a) for low temperatures where
        reproducible conductance fluctuations become visible.
     c) Conductance fluctuations $\delta G(B)$ after substraction of a smooth background
        $\overline{G(B)}$ for $B=6.97-7.02$~T and the same temperatures as in b).
        The scale of the plot is visualized by the grey
        area in b). The curves are shifted for clarity.}
  \label{data}
\end{figure}

Due to their low mobilities our samples show broad quantum Hall plateaus with
vanishing conductance $G=I/V$ around integer filling factors $\nu=nh/eB$. Near
half integer filling factors the conductance is nonzero with a peak value of
the order of $10\,e^2/h$, corresponding to diagonal conductivities
$\sigma_{xx}= (1/2\pi)\ln(1+w/r)G \approx 0.15\,e^2/h$, with an actual value
between 0.1 and 0.3$\,e^2/h$ (depending on sample and filling factor). Here we
concentrate on the plateau transition $\nu=2\rightarrow 1$ which is broadest in
magnetic field. The conductance at this transition is shown in Fig.~\ref{data}a
for different temperatures. It is dominated by the well understood conductance
peak which arises when the Fermi energy sweeps through the delocalized
electronic states near the center of a spin split Landau band.

Superimposed onto this conductance peak are small reproducible conductance
fluctuations, see Fig.~\ref{data}b. For further analysis they first have to be
separated from the peak form $\overline{G(B)}$ obtained by a polynomial
smoothing. Only features on scales $\Delta\!B > 50$~mT are included into
$\overline{G(B)}$. The final fluctuations $\delta G(B) = G(B) -
\overline{G(B)}$ extracted from the peak are shown in Fig.~\ref{data}c.
Distinct reproducible features can be clearly observed, they become smoother
with rising temperature and their amplitude shrinks.

Considering, that the electronic transport in the QH plateau transition with
extended states at the Fermi energy may be regarded as quasi metallic, it is
worthwhile to compare the observed conductance fluctuations with UCFs in
mesoscopic 2D metals, which are governed by the minimum of either the dephasing
length, $L_\phi=\sqrt{D\tau_\phi}$, or the thermal length, $L_T=\sqrt{\hbar
D/k_BT}$~\cite{lee87}. Here $D$ is the diffusion coefficient and $\tau_\phi$ is
the phase coherence time.

For the low temperatures  considered here ($T < 1$~K) it was shown
that in high magnetic fields electron-electron scattering dominates
$\tau_\phi$ in a 2DES realized in a semiconductor~\cite{brandes95}.
For a typical conductivity $\sigma_c\sim 0.5\,e^2/h$
near the critical point of the QH plateau transition it is estimated to
$\tau_\phi \sim\hbar/(k_BT)$~\cite{brandes94,polyakov98}. In this framework
the dephasing length $L_\phi$ approximately
equals the thermal length $L_T$ and a distinction between $L_\phi$ and $L_T$
becomes unnecessary.

The dephasing length $L_\phi(T)$ influences the temperature dependence of both
the amplitude of the conductance fluctuations, $\sqrt{\left<\delta
G^2\right>}$, and their correlations. This fact can be formalized when using a
temperature dependent correlation function averaged over the magnetic field,
$F(\Delta\!B) = \left<\delta G(B) \delta G(B+\Delta\!B)\right>_B$. Defining a
correlation field, $B_c(T) \approx\Phi_0/L_\phi(T)^2$, the correlation function
then embraces two limits~\cite{lee87}: It is constant for $\Delta\!B \ll B_c$,
i.e. $F(\Delta\!B) = F(0)=\left<\delta G^2\right>$, reduced to $F(B_c)=F(0)/2$
for $\Delta\!B = B_c$, and it follows $F(\Delta\!B) \propto 1/\Delta\!B^2$  for
$\Delta\!B \gg B_c$.

UCFs can be analyzed in terms of the power spectral density (PSD), defined as
the Fourier transform  of the correlation function $F(\Delta\!B)$
 \begin{equation} \mathcal{P}(f_B)=\frac{1}{2\pi}\int
    F(\Delta\!B)\exp({-i\,2\pi\,f_B\,\Delta\!B})\;d\Delta\!B
 \end{equation}
with $f_B$ the magnetic frequency. It has been shown~\cite{schaefer99} that
using the PSD yields a more accurate analysis of the data in an experimental
situation like ours, where we are only able to measure fluctuations on a
magnetic field scale $\Delta\!B < 0.05\,$T, and, where we are additionally
limited to a finite range of the magnetic field ($B=6.4-7.1$~T).

\begin{figure}
  \begin{center}
  \resizebox{0.9\plotwidth}{!}{\includegraphics{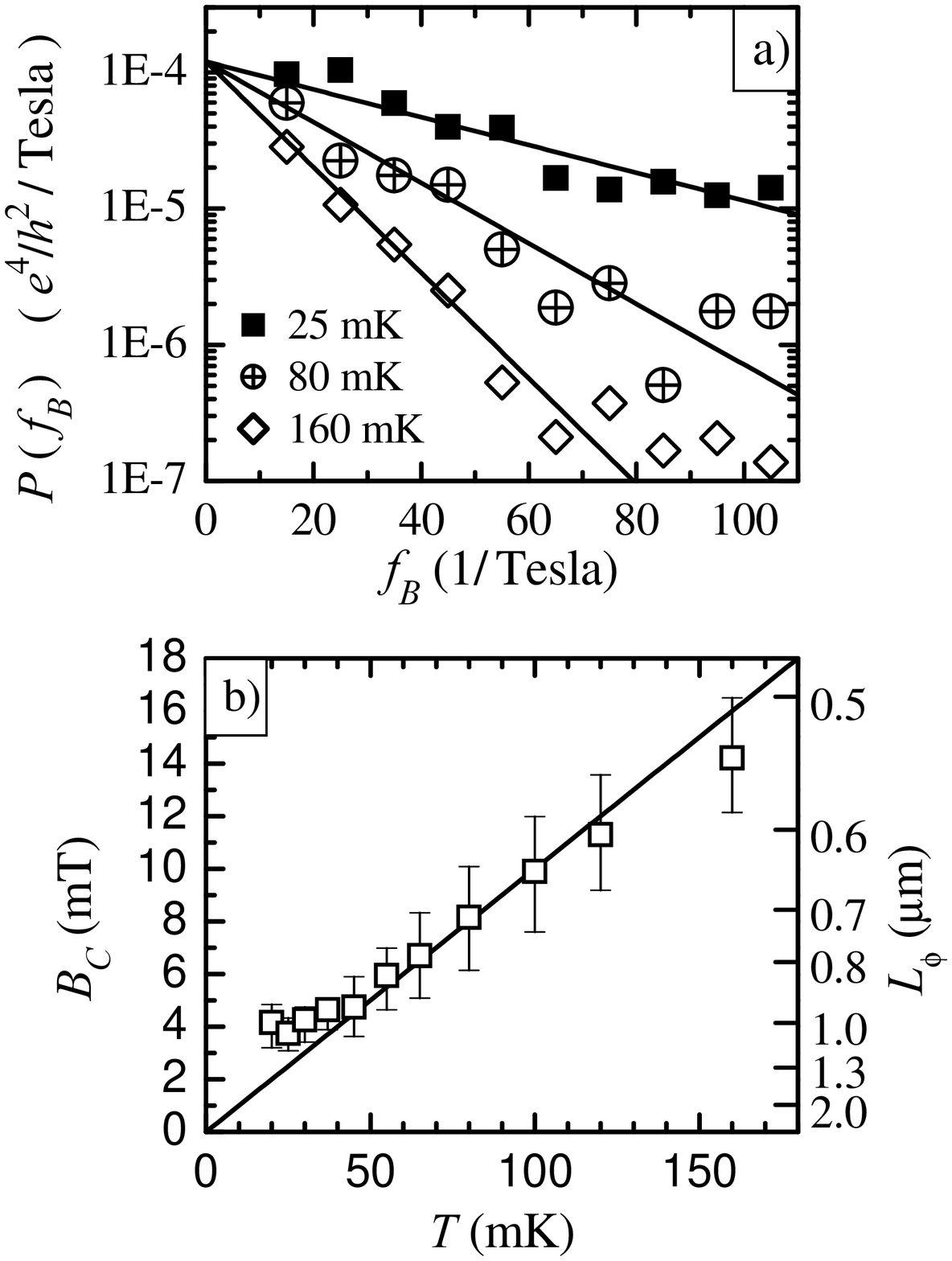}}
  \end{center}
  \caption{a) Power spectral density $\mathcal{P}(f_B)$
    of the conductance fluctuation $\delta G(B)$ in a magnetic field
    interval $B=6.4-7.1$~T. The lines show fits to Eq.~\ref{PSDeq}. \newline
    b) Correlation field $B_c$ resp. dephasing length $L_\phi=\sqrt{\Phi_0/B_c}$ determined
    by the fits shown in a). The straight line
    shows the expected linear dependence $B_c\propto T$ resp. $L_\phi \propto 1/\sqrt{T}$.
    }
  \label{PSD}
\end{figure}

In Fig.~\ref{PSD}a the PSD $\mathcal{P}(f_B)$ of our experimentally measured
conductance fluctuations, $\delta G(B) = G(B) - \overline{G(B)}$, is shown for
several temperatures. Each point within the graph is the result of averaging
$\ln(\mathcal{P}(f_B))$ over a frequency interval of $\Delta f_B=10$~T$^{-1}$.
The data follow an exponential decay indicated by the straight lines. The
observed saturation of the $T=160$~mK data for frequencies
$f_B\geq60\,$T$^{-1}$ is caused by measurement noise which is constant with
frequency and therefore always dominates the PSD at high $f_B$. For $T>200$~mK
the  frequency interval, where the PSD does not sink into the measurement
noise, becomes too small for a trustworthy analysis.

The exponential decay of the PSD
observed in Fig.~\ref{PSD}a can be fitted by a simple model
\begin{equation}
\mathcal{P}(f_B,B_c(T)) = \mathcal{P}_0\, e^{-2\pi B_c(T) f_B} \label{PSDeq}\,.
\end{equation}
Here $\mathcal{P}_0$ is a temperature independent prefactor. The corresponding
correlation function $F(\Delta\!B)$ fulfills all predictions for the limits of
$F(\Delta\!B)$. Additionally, the temperature dependence of $\left<\delta
G^2\right> = F(0)$ extracted from this PSD follows the expected behavior for a
2D metal, $\left<\delta G^2\right> \propto L_\phi^2(T) \propto
1/B_c(T)$~\cite{lee87}. A least square fit shown by the straight lines in
Fig.~\ref{PSD}a now enables us to determine the correlation field $B_c (T)$
plotted in Fig.~\ref{PSD}b. For $T\geq 50$~mK the correlation field follows a
linear dependence $B_c \propto T$ as expected for $B_c=\Phi_0/L_\phi^{2}$ and
$L_\phi\ \propto T^{-1/2}$. The corresponding temperature dependent dephasing
length $L_\phi$ is shown on the right axis of Fig.~\ref{PSD}b. Down to a
temperature $T = 50$ mK it is still well below the width of the sample, $w =
6\,\mu$m. The observed saturation of $L_\phi$ at temperatures below $T=50$~mK
is presumably caused by a decoupling of the electron temperature from the bath
temperature due to microwave heating.

Using the experimentally measured PSD we can now directly calculate the
amplitude square $\left<\delta G^2\right>$ of the UCFs. However, since our
procedure of extracting $\delta G$ from the peak disregards fluctuations on a
scale $\Delta\!B > 50\,$mT we have to restrict ourselves to frequencies $f_B$
above a cutoff frequency $f_g > (50\;{\rm mT})^{-1}$. This defines a reduced
fluctuation amplitude square
\begin{equation}
 \left<\delta G^2\right>_{\Delta\!B < (1/f_g)}=2\int_{f_g}^\infty
 df_B\,\mathcal{P}(f_B)
 \label{redAmp}
\end{equation}
where mainly fluctuations on small magnetic field scales $\Delta\!B < 1/f_g$
are taken into account. For a large enough $f_g$ effects of the smooth
background substraction are negligible and we can safely extract an
experimental UCF amplitude to be compared to theoretical calculations.

\begin{figure}
  \begin{center}
  \resizebox{0.85\plotwidth}{!}{\includegraphics{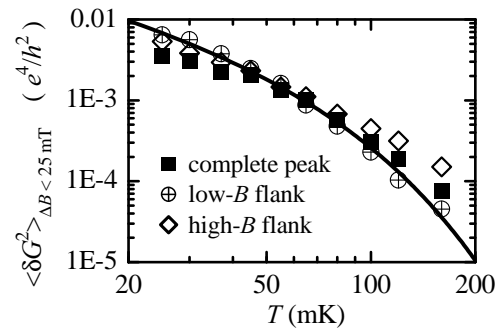}}
  \end{center}
  \caption{Reduced amplitude
    of the conductance fluctuations as defined in Eq.~\ref{redAmp} with
    $f_g=40$~T$^{-1}$. Data are shown for the complete magnetic field
    range $B=6.4-7.1\,$T used in Fig.~\ref{PSD}, for the low-$B$ flank in
    an interval $B=6.4-6.6\,$T, and the high-$B$ flank with $B=6.9-7.1\,$T.
    The line shows the expectation from Eqs.~(\ref{redAmp}) and (\ref{PSDeq})
    using $B_c ({\mathrm T}) = 0.1\cdot T ({\mathrm K})$
    linearly depending on $T$ as shown in Fig.~\ref{PSD}b.
}
  \label{Amp}
\end{figure}

Fig.~\ref{Amp} shows the measured reduced amplitude square of the fluctuations
as defined in Eq.~\ref{redAmp} for $f_g = 40\,$T$^{-1}$. The data are shown for
the complete magnetic field interval ({$6.4\;{\rm T} < B < 7.1\;$T})  and for
small $B$-intervals on both flanks of the peak around the points of half peak
height ({$6.4\;{\rm T} < B < 6.6\;$T and $6.9\;{\rm T} < B < 7.1\;$T}). The
experimental data are compared with the expectations from UCF theory, which
using Eqs.~(\ref{PSDeq}) and (\ref{redAmp}) yields \mbox{$\left<\delta
G^2\right>_{\Delta\!B < 1/f_g} = \mathcal{P}_0/(\pi B_c)\exp(-2\pi B_cf_g)$}
with a critical field $B_c$ linearly depending on $T$. As shown in
Fig.~\ref{Amp} the temperature dependence of all three data sets can be
described by UCF theory.

Using this reduced amplitude as given in Eq.~\ref{redAmp} or equivalent the
experimentally determined PSD as given in Eq.~\ref{PSDeq} we can extrapolate
from the raw experimental data to the full fluctuation amplitude
$\sqrt{\left<\delta G^2(T)\right>} \approx 0.02\;(e^2/h)\;(1/\sqrt{T({\rm K})})
\approx 0.1\;(e^2/h)\;L_\phi(\mu$m). This value can be compared to the
fluctuation amplitude
\begin{equation}
\sqrt{\left<\delta G^2(T)\right>} \sim \frac{e^2}{h} \left( \frac{2\pi r}{w}
\right)^{1/2} \frac{L_\phi (T)}{w}
\end{equation}
for a ring shaped metallic sample with radius $r$ and width $w$~\cite{lee87}.
An extension of UCF theory to the presence of Landau quantization does not
predict any change in the universal fluctuation
amplitude~\cite{xiong92,maslov93,khmelnitskii94,xiong97}. With the parameters
$w=6\,\mu$m and $r=60\,\mu$m for our sample we expect $\sqrt{\left< \delta
G^2\right>} \sim (e^2/h) L_\phi (\mu$m), one order of magnitude larger than the
experimental value deduced above. This discrepancy mirrors that the 2DES at the
QH plateau transition can be described to some extend like a 2D metal but is
not a metal. In fact, the conductivity $\sigma_{xx} < e^2/h$ does not allow the
perturbative treatment to first order of $(e^2/h)/\sigma_{xx}$ used in UCF
theory~\cite{lee87,xiong92} for metallic systems. However, numerical studies of
the two-point conductance $G$ of a rectangular sample at the transition from
the Hall insulator $G=0$ to the lowest Hall plateau $G=1$ with a Chalker and
Coddington network model~\cite{cho97,jovanovic98} also observe a reduced
amplitude $\sqrt{\left<\delta G^2(T)\right>} \lesssim 0.3e^2/h$ compared to
$\sqrt{\left<\delta G^2(T)\right>}\sim e^2/h$ for high filling.

Finally, the temperature dependence of the fluctuations amplitude can be used
to compare the length scale $L_\phi \sim L_T\propto T^{-1/2}$ governing the
fluctuations with the length $L_{sc}(T)\propto T^{-1/z}$ appearing in the
scaling behavior of the QH plateau transition, $z$ is the so-called dynamical
scaling exponent (for a review see~\cite{huckestein95}). For temperatures where
the sample width $w$ exceeds  $L_{sc}$  the transition width of the QH plateau
transition scales as $\Delta\!B \propto L_{sc}^{-1/\gamma} \propto T^\kappa$
with $\kappa=1/z\gamma$. For lower temperatures $\Delta\!B$ was experimentally
found to saturate~\cite{koch91}. This saturation could be attributed to finite
size scaling occuring for $w < L_{sc}$ which allowed a direct measurement of
the localization length in the QH plateau transition.

\begin{figure}
  \begin{center}
  \resizebox{\plotwidth}{!}{\includegraphics{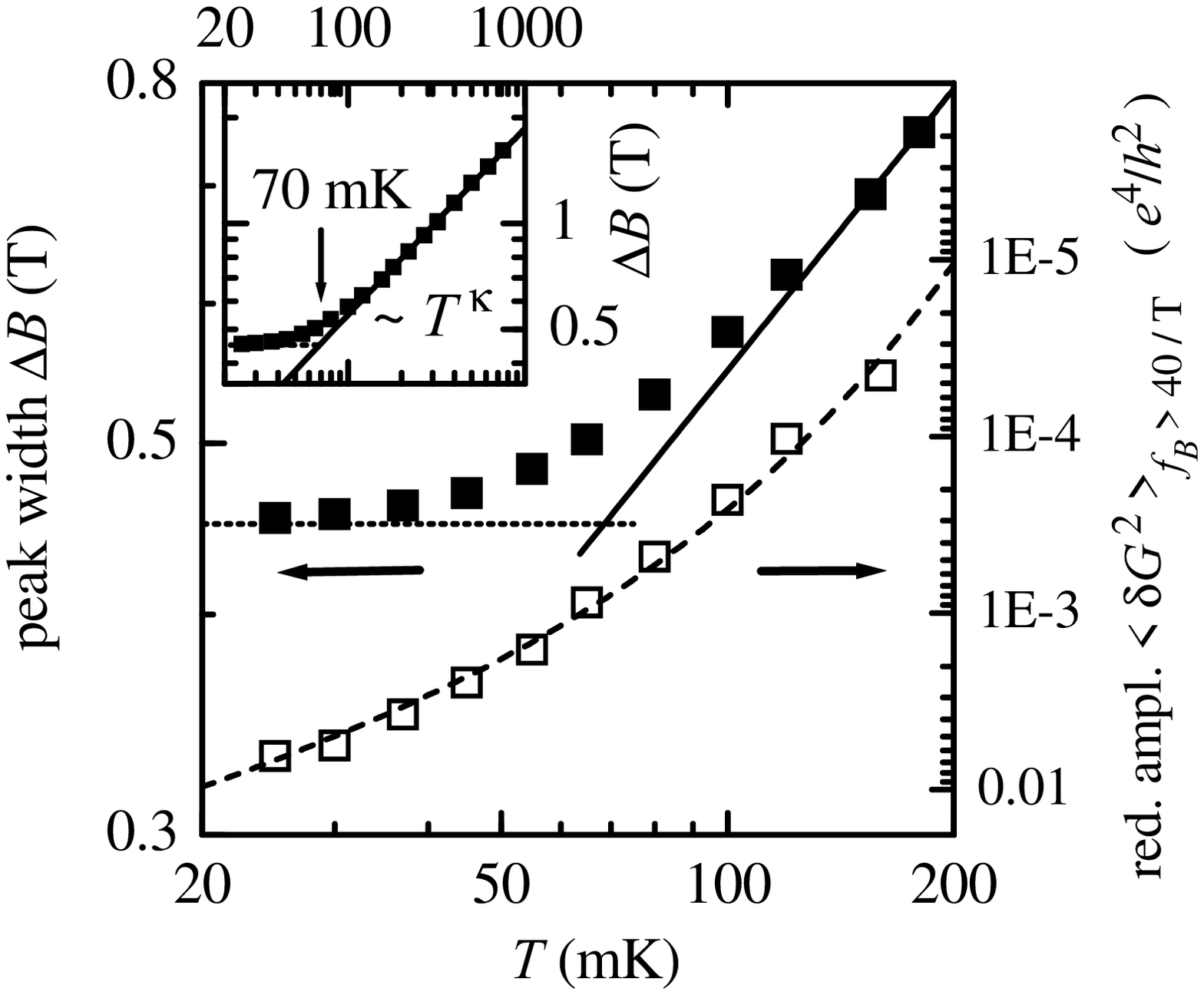}}
  \end{center}
  \caption{The peak width ($\blacksquare$), measured as full width at half maximum $\Delta\!B$,
    shows scaling behavior $\Delta\!B \propto T^\kappa$ (straight line) with
    $\kappa = 0.5$ at elevated temperatures (inset)
    and saturates due to finite size effects at low temperature
    (dotted line). A saturation of the electron temperature can be ruled out as the measured
    reduced fluctuation amplitude ($\square$, low-$B$-flank from
    Fig.~\ref{Amp}) follows the prediction (dashed line, $B_c ({\mathrm T}) = 0.1\cdot T ({\mathrm K})$)
    down to $T\leq 30$~mK.
}
  \label{width}
\end{figure}

Usually $L_{sc}$ is identified as the dephasing length $L_\phi$. However, it
was noted that for $L_\phi > w$ one would expect large mesoscopic
fluctuations~\cite{das-sarma94}. In contrast, no fluctuations~\cite{koch91} or
only small fluctuations~\cite{koch92} were observed experimentally at the
lowest temperatures. As a consequence, the observed saturation of $\Delta\!B$
was claimed to be caused by other effects such as external heating of the
2DES~\cite{das-sarma94}. In our experiments we can use the temperature
dependence of the fluctuation amplitude as a thermometer for the 2DES. As can
be seen from the temperature dependence of the correlation field in
Fig.~\ref{PSD}b and the fluctuation amplitude $\left<\delta G^2\right>$ in
Fig.~\ref{Amp} an electron temperature $T_e$ well below 50~mK is achieved. For
the lower conductance at the flanks, where unwanted microwave heating is
reduced, $\left<\delta G^2\right>$ even follows the model down 30~mK
(Figs.~\ref{Amp} and \ref{width}). In contrast, as shown in Fig.~\ref{width}
the width $\Delta\!B$ of the conductance peak only follows a scaling law
$\Delta\!B \propto T^\kappa$
for $T\geq150$~mK and saturates for $T<100$~mK. Interpreted as a
saturation decoupling of electron temperature from the temperature in
the mixing chamber one would find a minimum electron temperature $T_e=70$~mK
inconsistent with the observed temperature dependence of the fluctuations. This
demonstrates that the observed transition width saturation is indeed an effect
of the sample size. Additionally, using the generally accepted value
$\gamma=2.3$ of the critical scaling exponent and applying the relation
$\kappa = 1/z \gamma$ for the critical exponents of the QH plateau transition
one finds $L_{sc} \propto 1/T$~\cite{huckestein95}.
This deviates substantially from the temperature dependence of
$L_\phi \propto 1/\sqrt{T}$. Both observations demonstrate
that size scaling and conductance fluctuations are governed by distinct length
scales implying that $L_{sc}$ depends on different mechanisms than $L_\phi$,
an experimental finding in
agreement with recent theoretical predictions~\cite{polyakov98}.

In conclusion, we have measured the conductance fluctuations of a 2DES in the
quantum Hall regime. Their dependence on temperature and magnetic field can be
described by standard UCF-theory for metals, however, their amplitude is
considerably reduced. In addition we have shown that temperature dependence of
the quantum Hall plateau transition and that of the fluctuations are governed
by different length scales.

The samples used for our experiments were grown by K. Ploog at the Max Planck
Institut f\"ur Festk\"orperforschung. We thank F. Evers, F. Kuchar, D.
Polyakov, and L. Schweitzer for stimulating discussions.

\bibliographystyle{apsrev}

\end{document}